\documentclass[12pt]{article}
\usepackage{authblk}
\usepackage{geometry}
\usepackage{amsthm}
\usepackage[scr=esstix]{mathalpha}
\usepackage{caption}
\usepackage{subcaption}
\usepackage{customcommands}
\usepackage{bm}
\usepackage{Chrisstyle}
\usepackage{comment}	
\usepackage{todonotes}
\usepackage[export]{adjustbox}

\usepackage{amsmath}
\usepackage{amssymb}
\usepackage{amsfonts}
\usepackage{mathtools}
\usepackage{physics}
\usepackage{upgreek}
\usepackage{dsfont}
\usepackage[normalem]{ulem}
\usepackage{color}
\usepackage{bbold}
\usepackage{mathtools}
\usepackage{tensor} 
\usepackage{thmtools} 
\usepackage{thm-restate} 
\usepackage{subcaption}
\usepackage{braket}

\usepackage{tikz-cd}
\usepackage{tikz}
\usetikzlibrary{shapes.geometric}
\usetikzlibrary{calc}
\usetikzlibrary{arrows.meta}
\usetikzlibrary{decorations.pathreplacing}
\usepackage{array}

\usepackage{xcolor}
\colorlet{darkblue}{blue!70!black}
\colorlet{darkgreen}{green!50!black}
\colorlet{midgreen}{green!60!black}
\colorlet{darkbrown}{brown!70!black}

\definecolor{brownkenny}{HTML}{8B4513}
\definecolor{bluekenny}{HTML}{007Aff}
\definecolor{greenkenny}{HTML}{94E700}
\definecolor{orangekenny}{HTML}{FF8800}
\definecolor{redkenny}{HTML}{FF0000}
\definecolor{pinkkenny}{HTML}{FF9797}
\definecolor{darkbluekenny}{HTML}{03468F}
\definecolor{darkgreenkenny}{HTML}{007355}
\definecolor{purplekenny}{HTML}{82218b}

\newcommand\numberthis{\addtocounter{equation}{1}\tag{\theequation}}

\title{On observers in holographic maps}

\author{Chris Akers,}
\author{Gracemarie Bueller,}
\author{Oliver DeWolfe,}
\author{Kenneth Higginbotham,}
\author{\\ Johannes Reinking,}
\author{and Rudolph Rodriguez}

\affiliation{Department of Physics and\\ Center for Theory of Quantum Matter\\
University of Colorado\\ Boulder, CO 80309}

\emailAdd{chris.akers@colorado.edu}
\emailAdd{gracemarie.bueller@colorado.edu}
\emailAdd{oliver.dewolfe@colorado.edu}
\emailAdd{kenneth.higginbotham@colorado.edu}
\emailAdd{johannes.reinking@colorado.edu}
\emailAdd{rudolph.rodriguez@colorado.edu}

\abstract{
A straightforward gravitational path integral calculation implies that closed universes are trivial, described by a one dimensional Hilbert space. Two recent papers by Harlow-Usatyuk-Zhao and Abdalla-Antonini-Iliesiu-Levine have sought to ameliorate this issue by defining special rules to incorporate observers into the path integral. However, the proposed rules are different, leading to differing results for the Hilbert space dimension. Moreover, the former work offers a holographic map realized using a non-isometric code construction to complement their path integral result and clarify its physics. In this work, we propose a non-isometric code that implements the second construction, allowing thorough comparison. Our prescription may be thought of as simply removing the portion of the map that acts on the observer, while preserving the rest,  creating an effective holographic boundary at the observer-environment interface. This proposal can be directly applied to general holographic maps for both open and closed universes of any dimension.
}

\begin{document}
\maketitle

\section{Introduction}
\label{sec:intro}

The fact that holography predicts a trivial Hilbert space for closed universes is both well-documented and unsettling 
\cite{maldacena_wormholes_2004,
marolf_transcending_2020,
mcnamara_baby_2020,
antonini_holographic_2024,
usatyuk_closed_2024,
usatyuk_closed_2025}.
Prior work on effects due to the presence of an observer in a gravitating spacetime 
\cite{chandrasekaran_algebra_2023,
witten_algebras_2023,
witten_background-independent_2024,
vuyst_gravitational_2024,
kudler-flam_algebraic_2024,
chen_clock_2024,
kolchmeyer_chaos_2024,
maldacena_real_2024,
tietto_microscopic_2025} 
gives hope that including an observer in a closed universe could cure this issue. Recently, works by Harlow-Usatyuk-Zhao (HUZ) \cite{harlow_quantum_2025} and Abdalla-Antonini-Iliesiu-Levine (AAIL) \cite{abdalla_gravitational_2025} have proposed prescriptions for including an observer in the gravitational path integral in such a way as to produce a non-trivial fundamental Hilbert space. However, their prescriptions differ, resulting in two different dimensions for the fundamental Hilbert space of a closed universe.

The HUZ rules can be summarized as follows. Partition the Hilbert space of semiclassical gravity (the ``effective description'') into two factors,
\begin{equation}\label{eq:Heff}
    \mathcal{H}_\mathrm{eff} = \mathcal{H}_{Ob} \otimes \mathcal{H}_M
\end{equation}
where $\mathcal{H}_{Ob}$ is the Hilbert space of the observer and $\mathcal{H}_M$ that of matter in the observer's environment. Now say we consider two states $\ket{\psi}, \ket{\phi} \in \mathcal{H}_\mathrm{eff}$, and we wish to compute some moment $\lvert \braket{\psi|\phi} \rvert^m$ of their inner product using the gravitational path integral. Before doing the gravitational path integral, act on each state with an isometry
\begin{equation}
\begin{split}
    W:&~ \mathcal{H}_{Ob} \to \mathcal{H}_{Ob} \otimes \mathcal{H}_{Ob'}
\end{split}
\end{equation}
that clones the observer in the ``pointer'' basis of states stable under interactions with the environment, and \emph{treat that clone as a non-gravitational reference system} when doing the gravitational path integral. This has the effect of \emph{suppressing} certain terms in the path integral by a factor of order $1/d_{Ob} $, the dimension of the observer Hilbert space, see Figure \ref{fig:chutes&ladders}.

This suppression leads to a larger ``fundamental'' Hilbert space dimension computed by the path integral. A quick way to estimate this is via the variance in the fundamental inner product \cite{abdalla_gravitational_2025,akers_black_2024}. Imagine we have a Hilbert space $\mathcal{H}$ of unknown dimension, and we wish to estimate $ \dim \mathcal{H}$. All we are given is random vectors $\ket{i} \in \mathcal{H}$ for $i \in \{1, 2, ..., k\}$. It holds that
\begin{equation}\label{eq:variance_estimate}
     \sigma^2 = \overline{\lvert \braket{i|j} \rvert^2} - \Big|\overline{\braket{i|j}}\Big|^2 \approx \mathcal{O}\left( \frac{1}{\dim \mathcal{H}} \right),
\end{equation}
where overline denotes an average over different draws of $k$ random vectors. Hence by computing the variance in the inner product, we can estimate the Hilbert space's dimension. Similarly, we can use the gravitational path integral to learn the fundamental Hilbert space dimension by computing the first and second moments of the inner product and plugging them into \eqref{eq:variance_estimate}.

HUZ's observer rule suppresses\footnote{See for example equation (4.9) in \cite{harlow_quantum_2025}.} certain terms that show up in $\overline{\lvert \braket{i|j} \rvert^2}$ but not $\overline{\braket{i|j}}$, reducing the variance and thereby increasing $\dim  \mathcal{H}_\mathrm{fun}$. The suppressed terms can be characterized by the worldline of the observer. Consider the computation of 
\begin{equation}
    \lvert \braket{\psi|\phi} \rvert^2 = \lvert \braket{\psi|\phi}_1 \rvert \cdot \lvert \braket{\psi|\phi}_2 \rvert~.
\end{equation} 
Terms end up suppressed by order $1/d_{Ob}$ if the observer originating in ket 1 (respectively 2) does not terminate in bra 1 (respectively 2). For closed universes in JT gravity \cite{Jackiw:1984je, Teitelboim:1983ux}, this gives a fundamental Hilbert space of dimension
\begin{equation}\label{eq:HUY}
    \dim \mathcal{H}_\text{fun} \approx \min \left( d_{Ob}, e^{2S_0} \right)\,,
\end{equation}
which can be understood as saying the fundamental Hilbert space will be as large as the reference system ${\cal H}_{Ob'}$, unless there is a ``bottleneck" due to the finite size of the non-perturbative Hilbert space.

AAIL propose qualitatively different rules for including an observer. Rather than introduce a mechanism for suppressing terms, they simply \emph{discard} the terms in the path integral which are suppressed by HUZ, see Figure \ref{fig:chutes&ladders}. However, we are \emph{not} to interpret this as simply the $d_{Ob} \to \infty$ limit of the HUZ rule. Instead, the dimension computed with this rule is interpreted as the dimension of ``the Hilbert space relative to the observer,'' which we'll call $\mathcal{H}_\mathrm{rel}$. The \emph{total} dimension of the fundamental Hilbert space is the product of that with the observer dimension,
\begin{equation} \label{eq:AAIL}
    \dim \mathcal{H}_\text{fun} = \dim\left(\mathcal{H}_{Ob} \otimes \mathcal{H}_\mathrm{rel}\right) \approx d_{Ob} e^{2S_0}~.
\end{equation}
This differs substantially from \eqref{eq:HUY}, only agreeing in the limit that both $d_{Ob}$ and $S_0$ are taken to infinity or $d_{Ob} = e^{2 S_0} = 1$.

\begin{figure}
    \centering
    \begin{tabular}{>{\centering\arraybackslash}m{2.8cm}  >{\centering\arraybackslash}m{2.4cm}  >{\centering\arraybackslash}m{0.2cm} >{\centering\arraybackslash}m{2.1cm} >{\centering\arraybackslash}m{0.2cm} >{\centering\arraybackslash}m{2.1cm} > {\centering\arraybackslash}m{0.2cm} >{\centering\arraybackslash}m{0.8cm}}
    
        $\overline{|\langle\psi|V^\dagger V|\phi\rangle|^2} =$ &\begin{tikzpicture}
    \node (botr) at (0.75, -1.5) {}; 
    \node (topr) at (0.75, 1.5) {}; 
    \node (botl) at (-0.75, -1.5) {};
    \node (topl) at (-0.75, 1.5) {}; 

    \draw[fill=gray] (1.25,-1.5) -- (1.25, 1.5) -- (0.25,1.5) -- (0.25,-1.5) -- cycle;
    \draw[fill=gray] (-1.25,-1.5) -- (-1.25, 1.5) -- (-0.25,1.5) -- (-0.25,-1.5) -- cycle;

    \draw[fill=white] (botl) ellipse(5mm and 2mm);
    \draw[fill=white] (botr) ellipse(5mm and 2mm);
    \draw[fill=white] (topl) ellipse(5mm and 2mm);
    \draw[fill=white] (topr) ellipse(5mm and 2mm);
\end{tikzpicture}&+&\begin{tikzpicture}
    \node (botr) at (0.75, -1.5) {}; 
    \node (topr) at (0.75, 1.5) {}; 
    \node (botl) at (-0.75, -1.5) {};
    \node (topl) at (-0.75, 1.5) {}; 

    \draw[fill=gray] (1.25,-1.5) -- (-0.25, 1.5) -- (-1.25,1.5) -- (0.25,-1.5) -- cycle;
    \draw[fill=gray] (-1.25,-1.5) -- (0.25, 1.5) -- (1.25,1.5) -- (-0.25,-1.5) -- cycle;

    \draw[fill=white] (botl) ellipse(5mm and 2mm);
    \draw[fill=white] (botr) ellipse(5mm and 2mm);
    \draw[fill=white] (topl) ellipse(5mm and 2mm);
    \draw[fill=white] (topr) ellipse(5mm and 2mm);
\end{tikzpicture}&+&\begin{tikzpicture}
    \node (botr) at (0.75, -1.5) {}; 
    \node (topr) at (0.75, 1.5) {}; 
    \node (botl) at (-0.75, -1.5) {};
    \node (topl) at (-0.75, 1.5) {}; 

    \draw [fill=gray] (1.25,-1.5) arc(0:180:1.25) -- (-0.25,-1.5) arc (180:0:0.25)--cycle;
    \draw [fill=gray] (-1.25,1.5) arc(180:360:1.25) -- (0.25,1.5) arc (360:180:0.25)--cycle;

    \draw[fill=white] (botl) ellipse(5mm and 2mm);
    \draw[fill=white] (botr) ellipse(5mm and 2mm);
    \draw[fill=white] (topl) ellipse(5mm and 2mm);
    \draw[fill=white] (topr) ellipse(5mm and 2mm);
\end{tikzpicture} &+ & $\mathcal{O}(e^{-2S_0})$\\
        \vspace{0.5cm}\\
         & rigatoni &  & penne &  & macaroni&  & \\
        \vspace{0.5cm}\\
        Harlow et al.: & $\mathcal{O}\left(1\right)$ &  & $\mathcal{O}\left(\frac{1}{d_{Ob}}\right)$ & & $\mathcal{O}\left(\frac{1}{d_{Ob}}\right)$ & &\\
        \vspace{0.5cm}\\
        Abdalla et al.: & $\mathcal{O}\left(1\right)$ &  & 0 & & 0& &\\
    \end{tabular}
    \caption{Top row: leading contributions to the inner product squared from the gravitational path integral for a closed universe. All three terms are $\mathcal{O}(1)$. Middle row: by introducing an entangled non-gravitational copy of the observer, the work of HUZ suppresses the second and third terms by a factor of $1/d_{Ob}$. Bottom row: the work of AAIL insists that an observer must ``stay in their own universe'' and therefore removes the second and third terms from the path integral entirely.}
    \label{fig:chutes&ladders}
\end{figure}

Faced with these two different rules, we would like to more deeply understand the physics underlying each. One path is to understand a ``Hilbert space'' description of each rule, i.e. a description of the rule which does not reference the gravitational path integral (whose interpretation is not always manifest). HUZ provided such an understanding of their rule by constructing a \emph{holographic map with an observer} using a non-isometric code that reproduced their answer \eqref{eq:HUY}. A holographic map is simply a linear map from the effective Hilbert space to the fundamental Hilbert space
$$
    V: \mathcal{H}_\mathrm{eff} \to \mathcal{H}_\mathrm{fun}.
$$
When it happens that $\dim \mathcal{H}_\mathrm{eff} > \dim \mathcal{H}_\mathrm{fun}$, this holographic map is said to be a non-isometric code \cite{akers_black_2024}. The map tells us which states in the fundamental theory to associate to given states in the effective theory.\footnote{Studying such codes has been highly useful in understanding the quantum mechanics of black holes after the Page time \cite{akers_black_2024, kim_complementarity_2023,dewolfe_non-isometric_2023,dewolfe_bulk_2024, chandra_toward_2023,cao_overlapping_2025,bueller_tensor_2024, kar_non-isometric_2023, antonini_non-isometry_2024}.} Given two states $\ket{\phi},\ket{\psi} \in \mathcal{H}_\mathrm{eff}$ we can compute their ``fundamental inner product'' by mapping them both to the fundamental Hilbert space and taking the inner product there, $\braket{\phi| V^\dagger V |\psi}$. The gravitational path integral can be interpreted as giving statistical data about this fundamental inner product, so is ultimately determined by the underlying holographic map. HUZ demonstrate that the holographic map naturally associated to a closed universe has a one dimensional $\mathcal{H}_\mathrm{fun}$ due to the lack of a boundary, and furthermore show how to include an observer in a way that gives rise to their path integral rules and a larger $\mathcal{H}_\mathrm{fun}$.

So far, it has not been explained how to include an observer in a holographic map to give rise to the AAIL rule. Our main result is to close this gap, constructing such a map with observer, further clarifying the distinction between the two proposals. Our prescription can be summarized simply: we act only the part of the holographic map that does not act on the observer. This distinction requires a notion of locality in the effective description, but we believe this is a natural expectation. Under this modified map, the observer degrees of freedom persist into the fundamental Hilbert space, and the interface between the observer and her environment becomes a boundary on which additional degrees of freedom live. One way of understanding the triviality of closed universe holography without an observer is that there is no suitable spacetime boundary to contain the dual description; in our map, singling out an observer creates such a boundary, allowing the Hilbert space to be non-trivial.

Section \ref{sec:code} will describe our method of including an observer into the map. In order to allow a direct comparison with the work of HUZ, we start by implementing the map using a model similar to theirs, specifically the one found in Appendix B of \cite{harlow_quantum_2025}, and we explain how to change that map to obtain one implementing the AAIL rule. Then we generalize, explaining how the AAIL rule applies naturally in tensor network models. Section \ref{sec:comments} will conclude with some comments. The details of relevant calculations can be found in the appendices.

\section{Observer in the holographic map from locality}
\label{sec:code}

We now describe a prescription for incorporating an observer into a holographic map that utilizes locality to simply remove the observer from the map. This automatically preserves the observer in the fundamental Hilbert space, and generates an effective holographic boundary between the observer and their environment. Our construction also reproduces the contributions to inner products arising from the AAIL path integral prescription. We have three ingredients:
\begin{enumerate}
    \item The Hilbert space of the effective theory, $\mathcal{H}_\mathrm{eff}$,
    \item The Hilbert space of the fundamental theory, $\mathcal{H}_\mathrm{fun}$,
    \item The linear ``holographic'' map between them, $V: \mathcal{H}_\mathrm{eff} \to \mathcal{H}_\mathrm{fun}$.
\end{enumerate}
Since holographic maps have often been implemented in the language of error-correcting codes, we will often use the word ``code" interchangeably. 

In a model of a closed universe without an observer, $\mathcal{H}_\mathrm{fun}$ is one dimensional; we start by reviewing such a holographic map $V$. In order to make $\mathcal{H}_\mathrm{fun}$ non-trivial, one must define an observer-dependent map $V_{Ob}$. In HUZ’s approach, $V_{Ob}$ consists of cloning the observer’s Hilbert space, followed by applying $V$ unchanged. Here, our $V_{Ob}$ involves instead restricting $V$ so that it acts trivially on the observer and a patch of geometry around them. Crucially, this modification of $V$ gives exactly the result of the AAIL rule: when computing averages, we keep only the diagrams in which the observer ``stays in their own universe'', and interpret the resulting dimension as the \emph{relative} Hilbert space dimension.

\subsection{No observer}
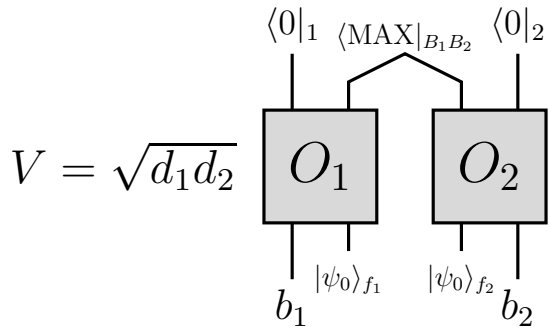
\begin{figure}
    \centering
    \scalebox{1.5}{
\begin{tikzpicture}

\node at (-1.25,0.5) {$V = \sqrt{d_1 d_2}$};

\node[scale=0.9] at (0.25,-0.75) {$b_1$};
\draw[thick] (0.25,-0.5) -- (0.25,0);
\node[scale=0.6] at (0.75,-0.5) {$|\psi_0\rangle_{f_1}$};
\draw[thick] (0.75,-0.25) -- (0.75,0);
\draw[fill=gray!30,thick] (0,0) rectangle (1,1);
\node[font=\large] at (0.5,0.5) {$O_1$};
\draw[thick] (0.25,1) -- (0.25,1.5);
\node[scale=0.75] at (0.25,1.75) {$\langle0|_1$};

\node[scale=0.9] at (2.25,-0.75) {$b_2$};
\draw[thick] (2.25,-0.5) -- (2.25,0);
\node[scale=0.6] at (1.75,-0.5) {$|\psi_0\rangle_{f_2}$};
\draw[thick] (1.75,-0.25) -- (1.75,0);
\draw[fill=gray!30,thick] (1.5,0) rectangle (2.5,1);
\node[font=\large] at (2,0.5) {$O_2$};
\draw[thick] (2.25,1) -- (2.25,1.5);
\node[scale=0.75] at (2.25,1.75) {$\langle0|_2$};

\draw[thick] (0.75,1) -- (0.75,1.25) -- (1.25,1.5) -- (1.75,1.25) -- (1.75,1);
\node[scale=0.6] at (1.25,1.65) {$\langle\text{MAX}|_{B_1 B_2}$};
    
\end{tikzpicture}
}
    \caption{The circuit diagram describing the model holographic map for a closed universe without an observer, given in (\ref{eq:Harlow}) and adapted from \cite{harlow_quantum_2025}.
    }
    \label{fig:Harlow}
\end{figure}

We first consider a non-isometric code for a closed universe without an observer. We will take HUZ's ``more structured code'' (see Appendix B of \cite{harlow_quantum_2025}) to be our holographic map $V$; see Fig.~\ref{fig:Harlow}. The effective Hilbert space is $\mathcal{H}_\mathrm{eff} = \mathcal{H}_{b_1} \otimes \mathcal{H}_{b_2}$, and we imagine that the ``bulk legs'' $b_1$ and $b_2$ are spatially separated from each other. A general map can be described in three steps: tensoring in additional Hilbert space factors, acting with a transformation on the enlarged space, and finally postselecting on some subspace.

First, we tensor in fixed states $|\psi_0\rangle_{f_1}$ and $|\psi_0\rangle_{f_2}$, and then we act random orthogonal matrices $O_1$ and $O_2$ on $b_1 f_1$ and $b_2 f_2$ respectively. Each orthogonal matrix is drawn independently from the Haar measure on the orthogonal group; the orthogonal group is chosen instead of the unitary group to be consistent with $\mathcal{CRT}$-invariance \cite{harlow_quantum_2025,harlow_gauging_2023}. 

Finally, we postselect on factors of the resulting Hilbert space.
Label the output Hilbert space of $O_1$ as $\mathcal{H}_1 \otimes \mathcal{H}_{B_1}$, and the output of $O_2$ as $\mathcal{H}_2 \otimes \mathcal{H}_{B_2}$, with the $\mathcal{H}_{B_i}$ having dimension $d_{B_1} = d_{B_2} = e^{2S_0}$, while the $\mathcal{H}_1$ and $\mathcal{H}_2$ represent everything else. The factors $B_1$ and $B_2$ are acted on by $\langle\text{MAX}|_{B_1 B_2}$, while the factors $\mathcal{H}_1$ and $\mathcal{H}_2$ are postselected on by $\langle0|_1$ and $\langle0|_2$ separately. We also multiply by $\sqrt{d_1 d_2}$, where 
\begin{equation}
    d_i \equiv \dim(\mathcal{H}_i \otimes \mathcal{H}_{B_i})
\end{equation}
in order to preserve normalization on average over the choice of $O_i$. 

Altogether, the holographic map described by Fig.~\ref{fig:Harlow} is given by 
\begin{equation} \label{eq:Harlow}
    V = \sqrt{d_1 d_2} \big( \langle0|_1 \langle0|_2 \langle\text{MAX}|_{B_1 B_2} \big) \big( O_1 \otimes O_2 \big) \big( \ket{\psi_0}_{f_1} \otimes \ket{\psi_0}_{f_2} \big),
\end{equation}
such that $V|\psi\rangle_{b_1 b_2} \in \mathcal{H}_\mathrm{fun}$.

\begin{figure}
    \centering
    \scalebox{0.85}{
\begin{tikzpicture}

\newcommand{\V}[5]{
    \begin{scope}[shift={(#1,#2)},yscale=#3]
        \draw[fill=gray!30,thick] (0,0) rectangle (1,1);
        \node[scale=1.4] at (0.5,0.5) {#4};
        \draw[fill=gray!30,thick] (1.5,0) rectangle (2.5,1);
        \node[scale=1.4] at (2,0.5) {#5};

        \draw[line width=2pt, redkenny] (0.25,-0.5) -- (0.25,0);
        \draw[thick] (0.75,-0.25) -- (0.75,0);
        \node[scale=0.6] at (0.75,-0.45) {$f_1$};
        \draw[thick] (1.75,-0.25) -- (1.75,0);
        \node[scale=0.6] at (1.75,-0.45) {$f_2$};
        \draw[line width=2pt, greenkenny] (2.25,-0.5) -- (2.25,0);

        \draw[thick] (0.25,1) -- (0.25,1.5);
        \draw[thick] (0.75,1) -- (0.75,1.25) -- (1.25,1.5) -- (1.75,1.25) -- (1.75,1);
        \draw[thick] (2.25,1) -- (2.25,1.5);
    \end{scope}
}

\newcommand{\RHS}[2]{
    \begin{scope}[shift={(#1,#2)}]
        \draw[line width=2pt, redkenny] (7.5,-0.5) -- (7.5,0);
        \draw[line width=2pt, redkenny] (7.5,3.5) -- (7.5,4);
        \draw[line width=2pt, greenkenny] (8,-0.5) -- (8,0);
        \draw[line width=2pt, greenkenny] (8,3.5) -- (8,4);
        \draw[decorate,decoration={brace,amplitude=6pt,mirror},thick] (7.4,-0.6) -- (8.1,-0.6);
        \node[scale=1] at (7.75,-1.15) {$|\phi\rangle$};
        \draw[decorate,decoration={brace,amplitude=6pt},thick] (7.4,4.1) -- (8.1,4.1);
        \node[scale=1] at (7.75,4.65) {$\langle\psi|$};
        
        \draw[line width=2pt, redkenny] (9,-0.5) -- (9,0);
        \draw[line width=2pt, redkenny] (9,3.5) -- (9,4);
        \draw[line width=2pt, greenkenny] (9.5,-0.5) -- (9.5,0);
        \draw[line width=2pt, greenkenny] (9.5,3.5) -- (9.5,4);
        \draw[decorate,decoration={brace,amplitude=6pt,mirror},thick] (8.9,-0.6) -- (9.6,-0.6);
        \node[scale=1] at (9.25,-1.15) {$|\psi\rangle$};
        \draw[decorate,decoration={brace,amplitude=6pt},thick] (8.9,4.1) -- (9.6,4.1);
        \node[scale=1] at (9.25,4.65) {$\langle\phi|$};
    \end{scope}
}


\V{0}{0}{1}{$O_1$}{$O_2$};
\node[scale=0.75] at (0.25,1.75) {$|0\rangle\langle0|$};
\node[scale=0.75] at (2.25,1.75) {$|0\rangle\langle0|$};
\V{0}{3.5}{-1}{$O_1^\intercal$}{$O_2^\intercal$};
\V{3}{0}{1}{$O_1$}{$O_2$};
\node[scale=0.75] at (3.25,1.75) {$|0\rangle\langle0|$};
\node[scale=0.75] at (5.25,1.75) {$|0\rangle\langle0|$};
\V{3}{3.5}{-1}{$O_1^\intercal$}{$O_2^\intercal$};

\draw[decorate,decoration={brace,amplitude=6pt,mirror},thick] (0,-0.6) -- (2.5,-0.6);
\node[scale=1] at (1.25,-1.15) {$|\phi\rangle$};
\draw[decorate,decoration={brace,amplitude=6pt},thick] (0,4.1) -- (2.5,4.1);
\node[scale=1] at (1.25,4.65) {$\langle\psi|$};
\draw[decorate,decoration={brace,amplitude=6pt,mirror},thick] (3,-0.6) -- (5.5,-0.6);
\node[scale=1] at (4.25,-1.15) {$|\psi\rangle$};
\draw[decorate,decoration={brace,amplitude=6pt},thick] (3,4.1) -- (5.5,4.1);
\node[scale=1] at (4.25,4.65) {$\langle\phi|$};

\draw[thick] [arrows = {-Latex[width=7pt, length=7pt]}] (6,1.75) -- (7,1.75);
\node[scale=0.8] at (6.5,2.15) {$\overline{O_1}\, \overline{O_2}$};

\RHS{0}{0};
\draw[thick] (7.5,0) -- (7.5,3.5);
\draw[thick] (8,0) -- (8,3.5);
\draw[thick] (9,0) -- (9,3.5);
\draw[thick] (9.5,0) -- (9.5,3.5);

\node[scale=1.25] at (10.5,1.75) {$+$};

\RHS{4}{0};
\draw[thick] (11.5,0) -- (13,3.5);
\draw[thick] (12,0) -- (13.5,3.5);
\draw[thick] (13,0) -- (11.5,3.5);
\draw[thick] (13.5,0) -- (12,3.5);

\node[scale=1.25] at (14.5,1.75) {$+$};

\RHS{8}{0};

\draw[thick] (15.5,0) -- (15.5,0.25) -- (17,0.25) -- (17,0);
\draw[thick] (16,0) -- (16,0.5) -- (17.5,0.5) -- (17.5,0);
\draw[thick] (15.5,3.5) -- (15.5,3.25) -- (17,3.25) -- (17,3.5);
\draw[thick] (16,3.5) -- (16,3) -- (17.5,3) -- (17.5,3.5);
    
\end{tikzpicture}
}
    \caption{A diagrammatic representation of $\overline{|\langle\psi|V^\dagger V|\phi\rangle|^2}$ (left side) and the three leading terms in the average over the Haar measure (right side). Red lines represent $b_1$ degrees of freedom, and green lines represent $b_2$ degrees of freedom. $f_i$ indicates the insertion of fixed states $|\psi_0\rangle_{f_i}$; these drop out in the average. Overall numerical prefactors have been omitted for convenience.}
    \label{fig:avg_Harlow}
\end{figure}
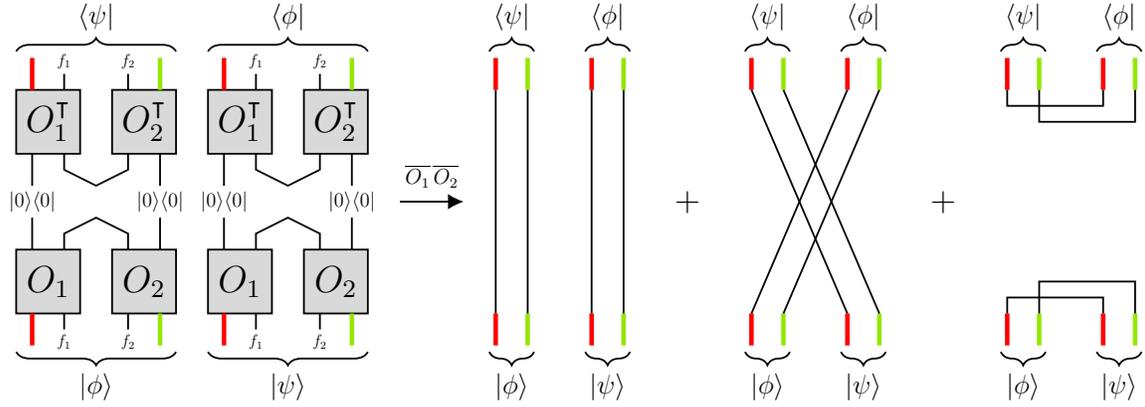

This $V$ postselects on the entire output, so manifestly $V$ maps $\mathcal{H}_\mathrm{eff}$ to a one-dimensional Hilbert space, as expected for a closed universe code. We could also see this from the large variance in inner products over the choice of $O_i$. Inner products of fundamental states can be computed on average using Haar measure integration tools; see Appendix A of \cite{harlow_quantum_2025} for averages over orthogonal matrices. It turns out that the averaged inner product on the fundamental Hilbert space is given by the inner product on the effective Hilbert space:
\begin{equation}
    \overline{\langle\psi|V^\dagger V|\phi\rangle} = \langle\psi|\phi\rangle.
\end{equation}
The average of the inner product squared has a total of 81 terms, only three of which are leading:
\begin{equation} \label{eq:Harlow_IP2_no_Obs}
    \overline{|\langle\psi|V^\dagger V|\phi\rangle|^2} = |\langle\psi|\phi\rangle|^2 + 1 + |\langle\psi^*|\phi\rangle|^2 + \mathcal{O}\left( e^{-2S_0} \right),
\end{equation}
where $|\psi^*\rangle$ denotes the $\mathcal{CRT}$-conjugate of $|\psi\rangle$. These leading terms are shown diagrammatically in Fig.~\ref{fig:avg_Harlow}, and are analogous to the geometries summed over in the path integral in Fig.~\ref{fig:chutes&ladders}.

The next subleading terms (of which there are six) are suppressed by $e^{-2S_0}$. The variance in the fundamental inner product is then
\begin{equation}
    \sigma^2 = \overline{|\langle\psi|V^\dagger V|\phi\rangle|^2} - \left\lvert\overline{ \langle\psi|V^\dagger V|\phi\rangle}\right\rvert^2 = 1 + |\langle\psi^*|\phi\rangle|^2 + \mathcal{O}\left( e^{-2S_0}\right).
\end{equation}
Hence from \eqref{eq:variance_estimate} we find the dimension of the fundamental Hilbert space is
\begin{equation}
    \dim \mathcal{H}_\text{fun} \approx \frac{1}{\sigma^2} = \mathcal{O}(1)~,
\end{equation}
again confirming $\mathcal{H}_\text{fun}$ is of order one.

\subsection{Including an observer}
Let us now assume that $b_1$ is occupied by an observer $Ob$. To incorporate this, we will define a variation of the holographic map, which we call $V_{Ob}$, that treats $b_1$ in a different way from the other bulk leg; $b_2$ we will treat the same as before, regarding it as occupied by some (non-observer) matter $M$. 

\begin{figure}
    \centering
    \scalebox{1.5}{
\begin{tikzpicture}

\node at (-1.25,0.5) {$V_{Ob} = \sqrt{\frac{d_2}{e^{2S_0}}}$};

\node[scale=0.9] at (0.5,-0.75) {$Ob$};
\draw[thick] (0.5,-0.5) -- (0.5,1.5);

\node[scale=0.9] at (2.25,-0.75) {$M$};
\draw[thick] (2.25,-0.5) -- (2.25,0);
\node[scale=0.6] at (1.75,-0.5) {$|\psi_0\rangle_{f_2}$};
\draw[thick] (1.75,-0.25) -- (1.75,0);
\draw[fill=gray!30,thick] (1.5,0) rectangle (2.5,1);
\node[font=\large] at (2,0.5) {$O_2$};
\draw[thick] (2.25,1) -- (2.25,1.5);
\node[scale=0.75] at (2.25,1.75) {$\langle0|_2$};
\draw[thick] (1.75,1) -- (1.75,1.5);
\node[scale=0.75] at (1.75,1.75) {$B_2$};
    
\end{tikzpicture}
}
    \caption{A circuit diagram describing the model holographic map for a closed universe with an observer, given by removing any operators and postselection acting on the observer.}
    \label{fig:Harlow_Obs}
\end{figure}
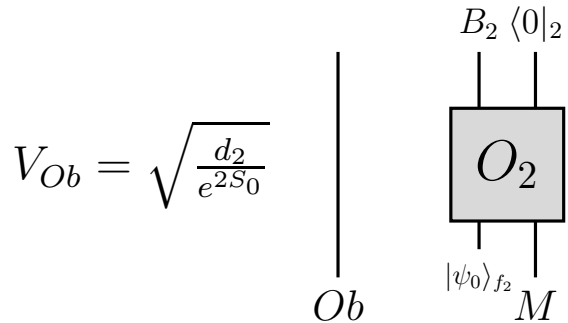

The new map $V_{Ob}$ is simply the old map $V$ except it does not include anything that acts on $b_1$. That is, we do not tensor in the fixed state $\ket{\psi_0}_{f_1}$, act with $O_1$, or postselect with $\bra{0}_1$ and $\bra{\mathrm{MAX}}_{B_1 B_2}$. The prefactor also changes: it is now just $\dim \mathcal{H}_2$, the dimension of the only thing being postselected on. The observer-included non-isometric code is then
\begin{equation} \label{eq:Harlow_Ob}
    V_{Ob} = \sqrt{\frac{d_2}{e^{2S_0}}} \big( \bra{0}_2 \big) \big( \mathbb{1}_1 \otimes O_2 \big) \big( \ket{\psi_0}_{f_2} \big) ~.
\end{equation}
See Figure \ref{fig:Harlow_Obs}.

The rationale is that this is a straightforward way to ensure the observer ends up as part of the fundamental Hilbert space. We just don't postselect on him! Notably, we are using the local structure of the holographic map, acting with ``just part'' of the map $V$ but not all of it. We will discuss the physical reasonableness of this in Section \ref{sec:comments}. For now, we emphasize that this does not mean we are removing the observer from the bulk. On the contrary, the bulk state and Hilbert space $\mathcal{H}_\mathrm{eff}$ are unchanged from before. Hence there is no concern that somehow the gravitational constraints might be violated. What has been modified is the map $V$ and the fundamental Hilbert space  $\mathcal{H}_\mathrm{fun}$. 

The dimension of the fundamental Hilbert space is now manifestly\footnote{The exact dimension will depend on details like the dimension of $\mathcal{H}_M$. The number quoted here assumes $\dim \mathcal{H}_M > e^{2S_0}$, which is true of the environment in the work of AAIL.} 
\begin{equation}
    \dim \mathcal{H}_\mathrm{fun} = \dim\left( \mathcal{H}_{Ob} \otimes \mathcal{H}_{B_2} \right) = d_{Ob} e^{2S_0}~.
\end{equation}
This can be seen from Figure \ref{fig:Harlow_Obs}: the output of $V_{Ob}$ consists of the legs $Ob$ and $B_2$.

\begin{figure}
    \centering
    \scalebox{0.85}{
\begin{tikzpicture}

\newcommand{\V}[4]{
    \begin{scope}[shift={(#1-0.25,#2)},yscale=#3]
        \draw[fill=gray!30,thick] (1.5,0) rectangle (2.5,1);
        \node[scale=1.4] at (2,0.5) {#4};

        \draw[line width=2pt, redkenny] (0.5,-0.5) -- (0.5,0);
        \draw[thick] (1.75,-0.25) -- (1.75,0);
        \node[scale=0.6] at (1.75,-0.45) {$f_2$};
        \draw[line width=2pt, greenkenny] (2.25,-0.5) -- (2.25,0);

        \draw[thick] (1.75,1) -- (1.75,1.5);
        \draw[thick] (2.25,1) -- (2.25,1.5);
    \end{scope}
}

\newcommand{\RHS}[2]{
    \begin{scope}[shift={(#1,#2)}]
        \draw[line width=2pt, redkenny] (7.5,-0.5) -- (7.5,0);
        \draw[line width=2pt, redkenny] (7.5,3.5) -- (7.5,4);
        \draw[line width=2pt, greenkenny] (8,-0.5) -- (8,0);
        \draw[line width=2pt, greenkenny] (8,3.5) -- (8,4);
        \draw[decorate,decoration={brace,amplitude=6pt,mirror},thick] (7.4,-0.6) -- (8.1,-0.6);
        \node[scale=1] at (7.75,-1.15) {$|\phi\rangle$};
        \draw[decorate,decoration={brace,amplitude=6pt},thick] (7.4,4.1) -- (8.1,4.1);
        \node[scale=1] at (7.75,4.65) {$\langle\psi|$};
        
        \draw[line width=2pt, redkenny] (9,-0.5) -- (9,0);
        \draw[line width=2pt, redkenny] (9,3.5) -- (9,4);
        \draw[line width=2pt, greenkenny] (9.5,-0.5) -- (9.5,0);
        \draw[line width=2pt, greenkenny] (9.5,3.5) -- (9.5,4);
        \draw[decorate,decoration={brace,amplitude=6pt,mirror},thick] (8.9,-0.6) -- (9.6,-0.6);
        \node[scale=1] at (9.25,-1.15) {$|\psi\rangle$};
        \draw[decorate,decoration={brace,amplitude=6pt},thick] (8.9,4.1) -- (9.6,4.1);
        \node[scale=1] at (9.25,4.65) {$\langle\phi|$};
    \end{scope}
}


\V{0}{0}{1}{$O_2$};
\draw[thick] (0.25,0) -- (0.25,3.5);
\draw[thick] (1.5,1.5) -- (1.5,2);
\node[scale=0.75] at (2,1.75) {$|0\rangle\langle0|$};
\V{0}{3.5}{-1}{$O_2^\intercal$};
\V{3}{0}{1}{$O_2$};
\draw[thick] (3.25,0) -- (3.25,3.5);
\draw[thick] (4.5,1.5) -- (4.5,2);
\node[scale=0.75] at (5,1.75) {$|0\rangle\langle0|$};
\V{3}{3.5}{-1}{$O_2^\intercal$};

\draw[decorate,decoration={brace,amplitude=6pt,mirror},thick] (0,-0.6) -- (2.25,-0.6);
\node[scale=1] at (1.125,-1.15) {$|\phi\rangle$};
\draw[decorate,decoration={brace,amplitude=6pt},thick] (0,4.1) -- (2.25,4.1);
\node[scale=1] at (1.125,4.65) {$\langle\psi|$};
\draw[decorate,decoration={brace,amplitude=6pt,mirror},thick] (3,-0.6) -- (5.25,-0.6);
\node[scale=1] at (4.125,-1.15) {$|\psi\rangle$};
\draw[decorate,decoration={brace,amplitude=6pt},thick] (3,4.1) -- (5.25,4.1);
\node[scale=1] at (4.125,4.65) {$\langle\phi|$};

\draw[thick] [arrows = {-Latex[width=7pt, length=7pt]}] (6,1.75) -- (7,1.75);
\node[scale=0.8] at (6.5,2.15) {$\overline{O_2}$};

\RHS{0}{0};
\draw[thick] (7.5,0) -- (7.5,3.5);
\draw[thick] (8,0) -- (8,3.5);
\draw[thick] (9,0) -- (9,3.5);
\draw[thick] (9.5,0) -- (9.5,3.5);

\node[scale=1.25] at (10.5,1.75) {$+\, e^{-2S_0}$};

\RHS{4}{0};
\draw[thick] (11.5,0) -- (11.5,3.5);
\draw[thick] (12,0) -- (13.5,3.5);
\draw[thick] (13,0) -- (13,3.5);
\draw[thick] (13.5,0) -- (12,3.5);

\node[scale=1.25] at (14.5,1.75) {$+\, e^{-2S_0}$};

\RHS{8}{0};

\draw[thick] (15.5,0) -- (15.5,3.5);
\draw[thick] (16,0) -- (16,0.5) -- (17.5,0.5) -- (17.5,0);
\draw[thick] (17,0) -- (17,3.5);
\draw[thick] (16,3.5) -- (16,3) -- (17.5,3) -- (17.5,3.5);
    
\end{tikzpicture}
}
    \caption{A diagrammatic representation of $\overline{|\langle\psi|V_{Ob}^\dagger V_{Ob}|\phi\rangle|^2}$ (left side) and the three leading terms in the average over the Haar measure for just $O_2$ (right side). Red lines represent observer $Ob$ degrees of freedom, and the green lines represent matter $M$ degrees of freedom. $f_i$ indicates the insertion of fixed states; these drop out in the average. Overall numerical prefactors have been omitted for convenience.}
    \label{fig:avg_Harlow_Obs}
\end{figure}
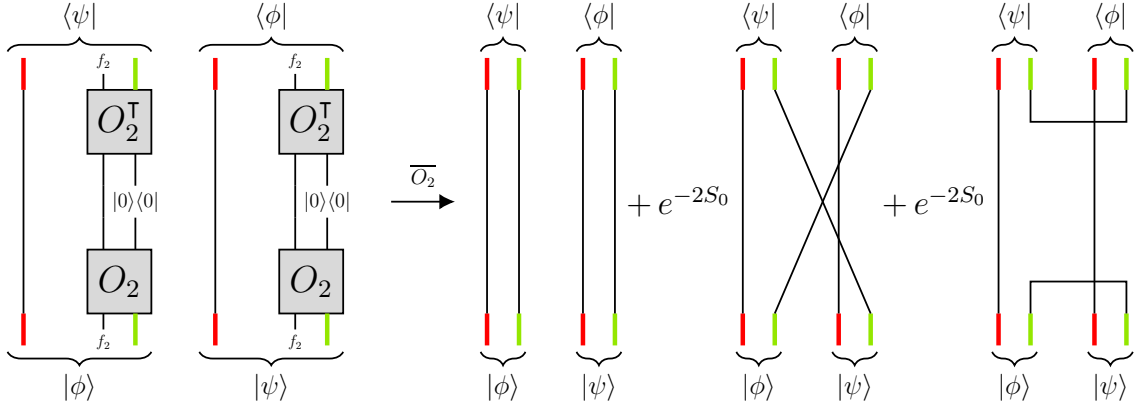

It is illuminating to also estimate this dimension from the variance in the inner product. This will illustrate that using $V_{Ob}$ implements the AAIL rule. We take $|\psi\rangle$ to be an arbitrary bipartite state on both $Ob$ and $M$ but will suppress these subscripts for ease of reading. With this observer-included map $V_{Ob}$, the averaged inner product remains unchanged,
\begin{equation} \label{eq:Harlow_IP1_w_Obs}
    \overline{\langle\psi|V_{Ob}^\dagger V_{Ob}|\phi\rangle} = \langle\psi|\phi\rangle,
\end{equation}
but the averaged inner product squared changes to
\begin{align*}
    \overline{|\langle\psi|V_{Ob}^\dagger V_{Ob}|\phi\rangle|^2} &= |\langle\phi|\psi\rangle|^2 \frac{d_2^2}{e^{2S_0}} (e^{2S_0} A + 2B) \\
        &\qquad + \Big( \tr \big[ \tr_M \rho_\psi \cdot \tr_M \rho_\phi \big] + \tr \big[ \rho_\psi^{\intercal_M} \cdot \rho_\phi^{\intercal_M} \big] \Big)  \frac{d_2^2}{e^{2S_0}} \big( A + B(e^{2S_0} + 1) \big) \numberthis \label{eq:Harlow_IP2_w_Obs}
\end{align*}
where
\begin{equation}
    A \equiv \frac{d_2 + 1}{d_2(d_2+2)(d_2-1)} \qquad B \equiv - \frac{1}{d_2(d_2+2)(d_2-1)}.
\end{equation}
We've denoted density matrices as $\rho_\psi \equiv |\psi\rangle\langle\psi|$, and $^{\intercal_M}$ denotes a partial transpose on $M$. To verify that this gives the same scaling of the Hilbert space dimension via the variance approximation, it is helpful to consider just the leading terms in (\ref{eq:Harlow_IP2_w_Obs}):
\begin{equation} \label{eq:Harlow_IP2_w_Obs_leading}
    \overline{|\langle\psi|V_{Ob}^\dagger V_{Ob}|\phi\rangle|^2} = |\langle\psi|\phi\rangle|^2 + e^{-2S_0} \Big( \tr \big[ \tr_M \rho_\psi \cdot \tr_M \rho_\phi \big] + \tr \big[ \rho_\psi^{\intercal_M} \cdot \rho_\phi^{\intercal_M} \big] \Big) + \mathcal{O}\left( \frac{1}{d_2} \right).
\end{equation}
These leading terms are shown diagrammatically in Fig.~\ref{fig:avg_Harlow_Obs}. Note how the observer $Ob$ (denoted by red in Fig.~\ref{fig:avg_Harlow_Obs}) is only acted on by the identity and thus always connects a ket to its matching bra. This is unlike the case of no observer in Fig.~\ref{fig:avg_Harlow}, where averaging over $O_1$ causes $b_1$ (also denoted by red) to connect with states in the other copy through the second and third terms. Therefore, our rule of removing $O_1$ when an observer is present implements AAIL's rule discarding contributions where the observer doesn't ``stay in their own universe.'' Furthermore, note that only the first term in Fig.~\ref{fig:avg_Harlow_Obs} is $\mathcal{O}(1)$. The other two are configurations that were present even in \eqref{eq:Harlow_IP2_no_Obs} but suppressed (like here) by $e^{-2 S_0}$, analogous to Fig.~\ref{fig:chutes&ladders}.

The leading $\mathcal{O}(1)$ term in (\ref{eq:Harlow_IP2_w_Obs_leading}) is precisely the term canceled by (\ref{eq:Harlow_IP1_w_Obs}) in the variation of the inner product. The next subleading terms are suppressed by a factor of $e^{2S_0}$; thus the variance of the inner product is on the order of $e^{-2S_0}$. We note that this does not capture the Hilbert space of the observer, which we've already taken to be a part of the fundamental Hilbert space and -- as seen in Fig.~\ref{fig:avg_Harlow_Obs} -- was not involved in the average. Therefore this variance estimates the size of the fundamental Hilbert space \textit{relative} to the observer:
\begin{equation} \label{eq:Hfun_w_obs}
    \mathcal{H}_\text{fun} = \mathcal{H}_{Ob} \otimes \mathcal{H}_\mathrm{rel}, \qquad \dim \mathcal{H}_\text{fun} \approx \frac{d_{Ob}}{\sigma^2} = d_{Ob} \, \mathcal{O}{\left( e^{2S_0} \right)}.
\end{equation}
This result precisely matches (\ref{eq:AAIL}) reported from AAIL's work. 

\subsection{More general effective Hilbert spaces} \label{sec:TN}
The codes $V$ and $V_{Ob}$ had very simple input Hilbert spaces, just two factors $\mathcal{H}_{b_1} \otimes \mathcal{H}_{b_2}$. A more realistic model would have many bulk inputs, $\mathcal{H}_{b_1} \otimes \mathcal{H}_{b_2} \otimes ... \otimes \mathcal{H}_{b_n}$. In this section we explain how such a model can be constructed, with observers easily included. The idea is simply that the holographic map is a \emph{tensor network}, like those of \cite{Swingle:2009bg, Pastawski:2015qua, Hayden:2016cfa}. Including an observer corresponds to removing one tensor from the network. (Again we emphasize that this does \emph{not} mean removing the observer from the bulk; the bulk Hilbert space is unchanged. ``Removing a tensor'' means changing the map $V$.)

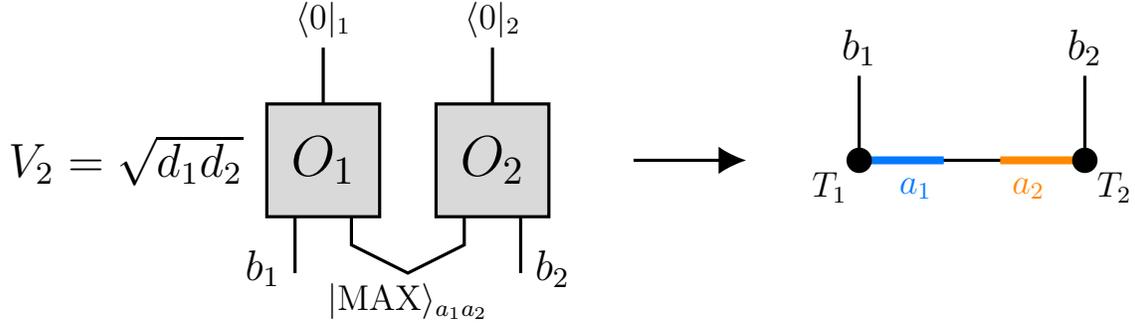
\begin{figure}
    \centering
    \scalebox{1.5}{
\begin{tikzpicture}

\node at (-1.25,0.5) {$V_2 = \sqrt{d_1 d_2}$};

\node[anchor=south east,scale=0.9] at (0.25,-0.75) {$b_1$};
\draw[thick] (0.25,-0.5) -- (0.25,0);
\draw[fill=gray!30,thick] (0,0) rectangle (1,1);
\node[font=\large] at (0.5,0.5) {$O_1$};
\draw[thick] (0.5,1) -- (0.5,1.5);
\node[scale=0.75] at (0.5,1.75) {$\langle0|_1$};

\node[anchor=south west,scale=0.9] at (2.25,-0.75) {$b_2$};
\draw[thick] (2.25,-0.5) -- (2.25,0);
\draw[fill=gray!30,thick] (1.5,0) rectangle (2.5,1);
\node[font=\large] at (2,0.5) {$O_2$};
\draw[thick] (2,1) -- (2,1.5);
\node[scale=0.75] at (2,1.75) {$\langle0|_2$};

\draw[thick] (0.75,0) -- (0.75,-0.25) -- (1.25,-0.5) -- (1.75,-0.25) -- (1.75,0);
\node[scale=0.75] at (1.25,-0.75) {$|\text{MAX}\rangle_{a_1 a_2}$};

\draw[thick] [arrows = {-Latex[width=7pt, length=7pt]}] (3.25,0.5) -- (4.25,0.5);

\draw[thick] (5.25,1.25) -- (5.25,0.5) -- (7.25,0.5) -- (7.25,1.25);
\draw[line width=2pt,bluekenny] (5.25,0.5) -- (6.,0.5);
\draw[line width=2pt,orangekenny] (6.5,0.5) -- (7.25,0.5);
\draw[fill=black] (5.25,0.5) circle (3pt);
\draw[fill=black] (7.25,0.5) circle (3pt);

\node[scale=0.75,anchor=north east] at (5.25,0.5) {$T_1$};
\node[scale=0.75,anchor=north west] at (7.25,0.5) {$T_2$};
\node[scale=0.9] at (5.25,1.5) {$b_1$};
\node[scale=0.9] at (7.25,1.5) {$b_2$};
\node[scale=0.75,bluekenny] at (5.75,0.25) {$a_1$};
\node[scale=0.75,orangekenny] at (6.75,0.25) {$a_2$};
    
\end{tikzpicture}
}
    \caption{A second holographic map $V_2$ without an observer, related to $V$ defined in Fig.~\ref{fig:Harlow} and equation (\ref{eq:Harlow}) by a partial transpose of $B_1B_2$. Two representations of $V_2$ are shown: as a circuit diagram (left) and as a 2-node tensor network (right).}
    \label{fig:SimpleTN}
\end{figure}

To start, note that the holographic map (without an observer) given in Fig.~\ref{fig:Harlow} and equation (\ref{eq:Harlow}) is related by a partial transpose to another holographic map where $|\text{MAX}\rangle$ is an input rather than a postselection; we define such a map $V_2$ as
\begin{equation}
    V_2 = \sqrt{d_1 d_2} \big( \langle0|_1 \langle0|_2 \big) \big( O_1 \otimes O_2 \big) |\text{MAX}\rangle_{a_1 a_2},
\end{equation}
where we've also relabeled the maximally entangled legs $B_1 B_2 \to a_1 a_2$ to emphasize this change. See the left side of Fig.~\ref{fig:SimpleTN} for a circuit representation of this holographic map. We note that both $O_i$ and the postselection $\langle0|_i$ now act on the bipartite system $b_i a_i$. We have removed the $f_i$ factors because they are not necessary, but they could still be included.

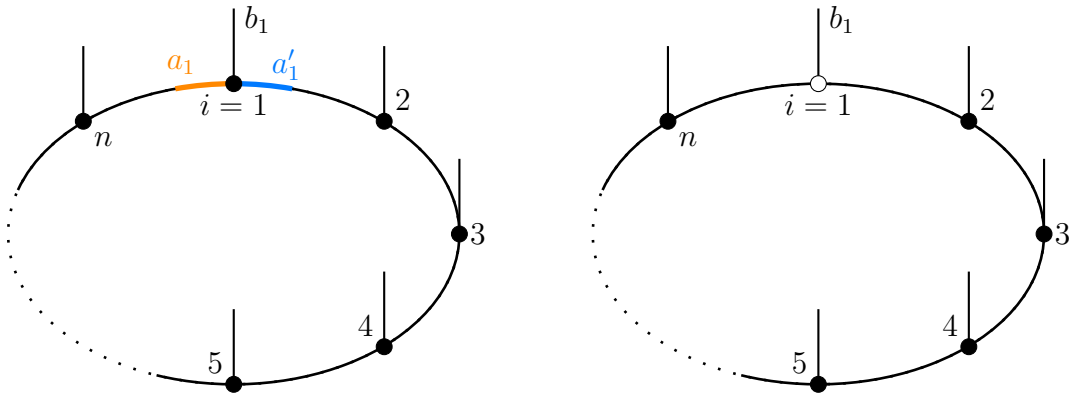
\begin{figure}
    \centering
    \begin{subfigure}{0.5\textwidth}
        \centering
            \begin{tikzpicture}
    \draw[line width = 1pt, dash pattern=on 1pt off 5pt] (3,0) arc [start angle = 0, end angle = 360, x radius = 3, y radius =2];
    
    \draw[line width = 1pt] (0,2) arc [start angle = 90, end angle = 163, x radius = 3, y radius = 2]; 
    \draw[line width = 1pt] (0,2) arc [start angle = 90, end angle = -110, x radius = 3, y radius = 2]; 

    \draw[line width = 2pt, bluekenny] (0,2) arc [start angle = 90, end angle = 75, x radius = 3, y radius = 2];
    \draw[line width = 2pt, orangekenny] (0,2) arc [start angle = 90, end angle = 105, x radius = 3, y radius = 2];

    \draw[fill] (2,1.5) circle (3pt);
    \draw[fill] (0,2) circle (3pt);
    \draw[fill] (-2,1.5) circle (3pt);
    \draw[fill] (0,-2) circle (3pt);
    \draw[fill] (2, -1.5) circle (3pt);
    \draw[fill] (3,0) circle (3pt);

    \draw[thick] (2,1.5) -- (2,2.5);
    \draw[thick] (0,2) -- (0,3);
    \draw[thick] (-2,1.5) -- (-2,2.5);
    \draw[thick] (0,-2) -- (0,-1);
    \draw[thick] (2, -1.5) -- (2, -0.5);
    \draw[thick] (3,0) -- (3,1);

    \node[anchor = south west] at (2,1.5) {$2$};
    \node[anchor = north] at (0,2) {$i=1$};
    \node[bluekenny] at (0.7,2.25) {$a'_1$};
    \node[orangekenny] at (-0.7,2.25) {$a_1$};
    \node[anchor = west] at (0,2.85) {$b_1$};
    \node[anchor = north west] at (-2, 1.5) {$n$};        
    \node[anchor = south east] at (0, -2) {$5$};
    \node[anchor = west] at (3,0) {$3$};
    \node[anchor = south east] at (2, -1.5) {$4$};

\end{tikzpicture}
    \end{subfigure}%
    \begin{subfigure}{0.5\textwidth}
        \centering
            \begin{tikzpicture}
    \draw[line width = 1pt, dash pattern=on 1pt off 5pt] (3,0) arc [start angle = 0, end angle = 360, x radius = 3, y radius =2];
    
    \draw[line width = 1pt] (0,2) arc [start angle = 90, end angle = 163, x radius = 3, y radius = 2]; 
    \draw[line width = 1pt] (0,2) arc [start angle = 90, end angle = -110, x radius = 3, y radius = 2]; 

    \draw[thick] (2,1.5) -- (2,2.5);
    \draw[thick] (0,2) -- (0,3);
    \draw[thick] (-2,1.5) -- (-2,2.5);
    \draw[thick] (0,-2) -- (0,-1);
    \draw[thick] (2, -1.5) -- (2, -0.5);
    \draw[thick] (3,0) -- (3,1);

    \draw[fill] (2,1.5) circle (3pt);
    \draw[fill=white] (0,2) circle (3pt);
    \draw[fill] (-2,1.5) circle (3pt);
    \draw[fill] (0,-2) circle (3pt);
    \draw[fill] (2, -1.5) circle (3pt);
    \draw[fill] (3,0) circle (3pt);

    \node[anchor = south west] at (2,1.5) {$2$};
    \node[anchor = north] at (0,2) {$i=1$};
    \node[anchor = west] at (0,2.85) {$b_1$};
    \node[anchor = north west] at (-2, 1.5) {$n$}; 
    \node[anchor = south east] at (0, -2) {$5$};
    \node[anchor = west] at (3,0) {$3$};
    \node[anchor = south east] at (2, -1.5) {$4$};

\end{tikzpicture}
    \end{subfigure}
    \caption{Pictorial representation of the $n$-node tensor network for a $(1+1)$-dimensional closed universe. Each node has three legs: one bulk leg $b_i$ and two ``in-plane legs'' $a_i$ and $a'_i$ connecting neighboring nodes. Left: no observer. Right: observer at node $i=1$; tensor $\langle T_1|$ has been removed, indicated by an open circle. Legs $b_1$, $a_1$, and $a'_1$ are all still present, but now considered to be ``boundary'' legs.}
    \label{fig:TN}
\end{figure}

Defining $\langle T_i|_{b_i a_i} \equiv \langle0|_i O_i$, we can rewrite $V_2$ in the more compact form
\begin{equation}
    V_2 = \sqrt{d_1 d_2} \big( \langle T_1|_{b_1 a_1} \otimes \langle T_2|_{b_2 a_2} \big) |\text{MAX}\rangle_{a_1 a_2}.
\end{equation}
In this form, we recognize $V_2$ as a 2-node tensor network, depicted graphically in the right hand side of Fig.~\ref{fig:SimpleTN}. Each $b_i$ represents a bulk leg, taking the bulk state $|\psi\rangle_{b_1 b_2} \in \mathcal{H}_\mathrm{eff}$ as inputs, while $a_i$ have become ``in-plane legs'' providing the spatial connectivity of the tensor network. These tensor networks provide a natural generalization of the previous subsection that allows for a more detailed description of the bulk locality. For example, we could consider an $n$-node random tensor network with periodic in-plane legs,
\begin{equation}
    |\text{MAX}\rangle_{a_1'a_2} |\text{MAX}\rangle_{a_2'a_3} \dots |\text{MAX}\rangle_{a_n'a_1},
\end{equation}
where link $a_i'a_{i+1}$ connects node $i$ to node $i+1$. This constructs a model holographic map for a $(1+1)$-dimensional closed universe with the topology of $S^1$:
\begin{equation} \label{eq:V_no_Obs}
    V_n = \left( \bigotimes_{i=1}^n \sqrt{d_i} \, \langle T_i|_{a_i b_i a'_i} \right) \left( \bigotimes_{i=1}^n |\text{MAX}\rangle_{a'_i a_{i+1}} \right).
\end{equation}
See the left panel of Fig.~\ref{fig:TN} for a pictorial representation of this tensor network. We now take $d_{a_i} = e^{S_0}$, different by a factor of $2$ in the exponent from the previous sections. (We can regard the previous codes as living on a very simple circle with two nodes, so $B_1 B_2$ was actually $a'_1 a_2$ \emph{and} $a'_2 a_1$, hence the squared dimension.) The factor $d_i = d_{a'_i}d_{a_i}d_{b_i}$ has been included to preserve normalization on average.

We implement the rules of the previous subsection for including an observer by removing the tensor at the node where the observer is located. For example, if an observer sits at node $i=1$, we remove $\langle T_1|$ so that the observer-included tensor network is given by
\begin{equation} \label{eq:V_w_Obs}
    V_{n,Ob} = \left( \bigotimes_{i=2}^n \sqrt{d_i} \, \langle T_i|_{a_i b_i a'_i} \right) \left( \bigotimes_{i=1}^n |\text{MAX}\rangle_{a'_i a_{i+1}} \right).
\end{equation}
Note that we have not removed either of the maximally entangled in-plane legs that end at node $i=1$ since they are connected to tensors at neighboring nodes. Pictorially, we denote an observer and its corresponding missing tensor as an open circle in the network, as shown in the right side of Fig.~\ref{fig:TN}.

Computations of the averaged inner product and its variance proceed similarly to the previous subsection (though are in fact much simpler using random tensor network techniques). Further details, including new rules for random tensor networks with orthogonal matrices, can be found in Appendix \ref{app:orthogonal}. In summary, these tensor networks give the same approximate fundamental Hilbert space dimensions:
\begin{align*}
    \text{No observer:}& \qquad \dim \mathcal{H}_\mathrm{fun} = 1 \\
    \text{Observer:}& \qquad \dim \mathcal{H}_\mathrm{fun} = d_{Ob}e^{2S_0}
\end{align*}
These results can be directly seen from Fig.~\ref{fig:TN}, or argued via a Renyi entropy calculation: see Appendix \ref{app:renyi}. Without an observer (left panel), there are no boundary legs, and thus the fundamental Hilbert space is trivial. With an observer (right panel), the $b_1$, $a_1$, and $a'_1$ legs are dangling since $\bra{T_1}$ is no longer present. These can be interpreted as boundary degrees of freedom; each of the two in-plane legs contributes $e^{S_0}$ to the fundamental Hilbert space dimension, so the total dimension is $d_{Ob}e^{2S_0}$.

\section{Comments}
\label{sec:comments}

We have introduced a new method for including observers in holographic codes. We simply act only the part of the holographic map that does not act on the observer. In both the circuit model of Fig.~\ref{fig:Harlow_Obs} and the tensor network model of Fig.~\ref{fig:TN}, this amounts to removing any operators, postselection, or tensors applied to the observer. This implements the rule of AAIL by removing the observer from the effects of averaging, eliminating any diagrams in computations that involve the observer not ``staying in their own universe.'' This provides a Hilbert space understanding of the AAIL rule, and makes manifest certain aspects, like that the fundamental Hilbert space dimension is
\begin{equation} \label{eq:general_Hfun}
    \dim \mathcal{H}_\mathrm{fun} = d_{Ob} e^{A/4G}
\end{equation}
where $G$ is Newton's constant and $A$ is the area of the surface surrounding the patch of spacetime containing the observer. In JT gravity, that $A/4G$ becomes $2S_0$.

\subsection*{Factorizability of the holographic map}
To include an observer as we do, it is crucial that the holographic map has a spatially local structure, similar to how a tensor network involves many little $O_i$ rather than one big $O$ acting non-locally on all bulk inputs. Only then can it be well-defined to act only part of the map (for us, the part that does not act on the observer). We do not view it as a shortcoming to have this local structure in the map. We see it as highly expected.

For example, the quantum extremal surface formula \cite{Engelhardt:2014gca} suggests the holographic map is spatially local in this way \cite{Akers:2021fut, Akers:2023fqr}. In real holographic maps, the entanglement wedge of a boundary region is found by extremizing over \emph{the spatial position} of the surface. In these model codes with random $O_i$, a quantum extremal surface prescription can also be derived, where the extremization is over \emph{which $O_i$'s inputs are included} in the entanglement wedge. This strongly suggests that real holographic maps are analogous to the model codes with many $O_i$ all acting in a spatially local way. 

This factorizability also plays nicely with a number of existing ideas, such as the ``generalized entanglement wedges'' of Bousso and Penington \cite{bousso_entanglement_2023, Bousso:2023sya}. Those authors prescribe a way to assign an entanglement wedge to any (gauge-invariantly defined) gravitational subregion, generalizing the idea of entanglement wedges for CFT subregions in AdS/CFT. This presupposes the same principle as us, that it is sensible to treat a given gravitational subregion like part of the fundamental Hilbert space, acting the holographic map only on everything else. We regard this principle as supported by the self-consistency of their proposal, which they demonstrate.

\subsection*{Comparison with HUZ}
Our explicit use of the factorizability of the holographic map underpins the difference between our code construction and that of HUZ. Both constructions find some way to ensure the observer ends up as part of the fundamental Hilbert space. In the HUZ construction, this is accomplished by cloning the observer. The holographic map is unchanged, so it does not act on this new clone factor. This is arguably the most natural way to treat the observer differently from other matter, in a way that ensures some version of the observer ends up as part of $\mathcal{H}_\mathrm{fun}$, if you don't want to use the spatial locality of the holographic map. The result is that the observer Hilbert space (and nothing additional) becomes the fundamental Hilbert space,
\begin{equation}
    \text{HUZ:}~~~\mathcal{H}_\mathrm{fun} \subseteq \mathcal{H}_{Ob}~,
\end{equation}
with $\mathcal{H}_\mathrm{fun} \subset \mathcal{H}_{Ob}$ in the case that $d_{Ob} > e^{2S_0}$. 

In our model, the fundamental Hilbert space includes the observer \emph{along with} the patch of geometry immediately around her. As a result, there is a stark difference in where information is located in the fundamental description. In that of HUZ, \emph{all} information ends up in the observer's Hilbert space. In our model, if $e^{2S_0} \gg d_{Ob}$, then almost all of the information about the environment ends up in $\mathcal{H}_\mathrm{rel}$. Very little ends up in the observer's factor (and none at all, if the observer is unentangled with the environment in the effective description).  

We summarize the difference as follows: The holographic map of HUZ holographically encodes all of $\mathcal{H}_\mathrm{eff}$ into $\mathcal{H}_{Ob}$. Our map picks a patch of geometry that includes the observer, and holographically encodes all of $\mathcal{H}_\mathrm{eff}$ outside that patch into the boundary of that patch plus the degrees of freedom inside.

\subsection*{Black holes}
Interestingly, there are many similarities between black holes and our method for including observers. In tensor networks, black holes are often represented by excising tensors in the black hole's interior, leaving dangling legs across the horizon \cite{Pastawski:2015qua, Hayden:2016cfa}. These additional horizon legs are exactly what give rise to the usual Bekenstein-Hawking entropy $A/4G$. In the same way, removing tensors at an observer's location leaves dangling legs at the boundary of the observer that contribute exactly the same kind of area term to the fundamental Hilbert space (\ref{eq:general_Hfun}). 

However, there are a few key differences between black holes and observers in our work. First, dangling horizon legs are taken to be inputs in the HaPPy code \cite{Pastawski:2015qua}, mapping the black hole's microstate to the boundary. Here, we take the observer's boundary legs to be part of the fundamental description, and therefore they act as outputs of the tensor network. Second, when the black hole interior is excised from the network, the interior bulk legs are removed as well; they are neither mapped to the boundary nor preserved as part of the fundamental Hilbert space. This is not so with observers in this work, whose bulk legs are retained as a part of the fundamental Hilbert space. 

It would be interesting to further study the effects of an observer \textit{inside} a black hole by using the methods of this work to include an observer in tensor networks for black hole interiors \cite{bueller_tensor_2024}.

\subsection*{Disjoint and extended observers}
Throughout this work, we have assumed that $(i)$ there is only one observer in the universe and $(ii)$ they occupy no more or less than one node in the tensor network. However, our construction is not limited to this assumption. For example, one could include multiple observers at various different nodes in the network by removing each of the tensors at those nodes.

Furthermore, it is possible to include a single ``large'' observer that takes up multiple neighboring nodes in the tensor network. In this case, all tensors in this neighborhood should be removed, along with all in-plane legs entirely contained within the neighborhood. This will result in the observer persisting in the fundamental Hilbert space, without introducing any spurious degrees of freedom. In-plane legs straddling the interface between observer and environment will be kept, and will produce the ${\cal H}_{\rm rel}$ degrees of freedom.

Finally, we note that this method cannot distinguish between one large observer and multiple independent observers located at neighboring nodes. We leave the philosophical implications of this lack of distinguishability to the reader.

\section*{Acknowledgments}

We are grateful to Daniel Harlow, Luca Iliesiu, Adam Levine, Mudassir Moosa, Mykhaylo Usatyuk, and Ying Zhao for helpful discussions. The authors are supported by the Department of Energy under grant DE-SC0010005 and by the Heising-Simons Foundation under grant 2024-4848. RR is supported by NSF grant HRD 2204630 via a fellowship.

\appendix
\section{Tensor networks with orthogonal matrices}
\label{app:orthogonal}

In this appendix we introduce the necessary tools for computing averaged inner products using random tensor networks with orthogonal matrices in Section \ref{sec:TN}. These rely on averages over the Haar measure on the orthogonal group; see Appendix A of \cite{harlow_quantum_2025} for further details. We use the abbreviated notation $|T_i\rangle_{a_i b_i a'_i} \equiv |T_i\rangle$. 

The averaged fundamental inner product can be written as
\begin{equation} \label{eq:avg_IP}
    \overline{\langle\psi|V_n^\dagger V_n|\phi\rangle} = \langle\tilde\psi| \left( \bigotimes_{i=1}^n d_i \, \overline{|T_i\rangle\langle T_i|} \right) |\tilde\phi\rangle,
\end{equation}
where we've defined
\begin{equation}
    |\tilde\psi\rangle = |\psi\rangle_{b_1 b_2 \ldots b_n} \bigotimes_{i=1}^n |\text{MAX}\rangle_{a'_i a_{i+1}},
\end{equation}
which involves the second moment of the Haar measure:
\begin{equation}
    \overline{|T_i\rangle\langle T_i|} \equiv \int dO \, O^\intercal |0\rangle\langle 0| O =  \mathbb{1}_i / d_i~,
\end{equation}
where $d_i \equiv d_{a'_i}d_{b_i}d_{a_{i+1}}$. Inserting this into (\ref{eq:avg_IP}) gives
\begin{equation}
    \overline{\langle\psi|V_n^\dagger V_n|\phi\rangle} = \langle\tilde\psi|\tilde\phi\rangle = \langle\psi|\phi\rangle.
\end{equation}
The result is unchanged if we replace $V_n$ with $V_{n,Ob}$.

The average inner product squared is given by
\begin{equation}
    \overline{|\langle\psi|V_n^\dagger V_n|\phi\rangle|^2} = \tr \left[ \Big( |\tilde\psi\rangle\langle\tilde\phi| \otimes |\tilde\phi\rangle\langle\tilde\psi| \Big) \Big( \bigotimes_i d_i^2 \, \overline{|T_i\rangle\langle T_i| \otimes |T_i\rangle\langle T_i|} \Big) \right]
\end{equation}
This involves the fourth moment of the Haar measure:
\begin{equation}
    \overline{|T_i\rangle\langle T_i| \otimes |T_i\rangle\langle T_i|} \equiv \int dO \, ( O^\intercal \otimes O^\intercal ) \, |0\rangle |0\rangle \langle0| \langle0| \, ( O \otimes O ).
\end{equation}
Normally, the fourth moment of the Haar measure on orthogonal matrices would involve nine terms; see Fig.~25 of \cite{harlow_quantum_2025}. Here, the factor of $|0\rangle |0\rangle \langle0| \langle0|$ sandwiched between the four copies of $O$ reduces the average to three terms:
\begin{equation}
    \int dO 
    \vcenter{
    \hbox{
    \scalebox{0.7}{
\begin{tikzpicture}

\newcommand{\Obox}[3]{
    \begin{scope}[shift={(#1,#2)}]
        \draw[fill=gray!30,thick] (0,0) rectangle (1,1);
        \node[font=\Large] at (0.5,0.5) {#3};
        \draw[thick] (0.5,-0.5) -- (0.5,0);
        \draw[thick] (0.5,1) -- (0.5,1.5);
    \end{scope}
}

\Obox{0}{0}{$O$};
\node[font=\large] at (0.5,2) {$|0\rangle\langle0|$};
\Obox{0}{3}{$O^\intercal$};

\Obox{1.5}{0}{$O$};
\node[font=\large] at (2,2) {$|0\rangle\langle0|$};
\Obox{1.5}{3}{$O^\intercal$};

\end{tikzpicture}
}
    }
    }
    = \frac{1}{d(d+2)} \Bigg(  
    \vcenter{
    \hbox{
    \scalebox{0.7}{
\begin{tikzpicture}

\draw[thick] (0,0) -- (0,2);
\draw[thick] (1,0) -- (1,2);

\node[font=\large] at (1.5,1) {$+$};

\draw[thick] (2,0) -- (2,0.5) -- (3,1.5) -- (3,2);
\draw[thick] (3,0) -- (3,0.5) -- (2,1.5) -- (2,2);

\node[font=\large] at (3.5,1) {$+$};

\draw[thick] (4,0) -- (4,0.5) -- (5,0.5) -- (5,0);
\draw[thick] (4,2) -- (4,1.5) -- (5,1.5) -- (5,2);

\end{tikzpicture}
}
    }
    }
    \Bigg).
\end{equation}
(Here $d$ is a stand-in for an arbitrary $d_i$.)
These three terms look schematically like the three leading terms in the gravitational path integral depicted in Fig.~\ref{fig:chutes&ladders}, and we will express them as $\mathbb{1}$, $F$, and $M$ respectively. Here $F$ is the swap operator, acting as $F \ket{a}_1 \ket{b}_2 = \ket{b}_1 \ket{a}_2$, and $M \equiv d|\text{MAX}\rangle\langle\text{MAX}|$.\footnote{We note that a Haar average over the unitary group would not have included the third term, $M \equiv d|\text{MAX}\rangle\langle\text{MAX}|$. While the $F$ term alone is enough to reproduce the trivial Hilbert space without an observer, the $M$ term is also needed to reproduce the results of the gravitational path integral.} Hence,
\begin{equation}
\label{eq:operator-products}
    \overline{|\langle\psi|V_n^\dagger V_n|\phi\rangle|^2} = \tr \left[ \Big( |\tilde\psi\rangle\langle\tilde\phi| \otimes |\tilde\phi\rangle\langle\tilde\psi| \Big) \Big( \bigotimes_i \frac{d_i^2}{d_i(d_i+2)} (\mathbb{1}_i + F_i + M_i) \Big) \right].
\end{equation}
The leading order contributions to this average are given by terms where each node is acted on by the same operator: $\mathbb{1}$, $F$, or $M$. Terms where nodes are acted on by some combination of the three operators are suppressed by (at least) $\mathcal{O}(d_a^{-1})$. When no observer is present, three terms dominate:
\begin{align*}
    \overline{|\langle\psi|V_n^\dagger V_n|\phi\rangle|^2} &= \tr \left[ |\tilde\psi\rangle\langle\tilde\phi| \otimes |\tilde\phi\rangle\langle\tilde\psi| \right] + \tr \left[ (\otimes_i F_i) \big( |\tilde\psi\rangle\langle\tilde\phi| \otimes |\tilde\phi\rangle\langle\tilde\psi| \big) \right] \\
        &\qquad + \tr\left[ (\otimes_i M_i) \big( |\tilde\psi\rangle\langle\tilde\phi| \otimes |\tilde\phi\rangle\langle\tilde\psi| \big) \right] + \mathcal{O}\left( \frac{1}{d_a^2} \right) \\
        &= |\langle\psi|\phi\rangle|^2 + 1 + |\langle\psi^*|\phi\rangle|^2 + \mathcal{O}\left( \frac{1}{d_a^2} \right). \numberthis
\end{align*}
Note that these are the same leading terms found in (\ref{eq:Harlow_IP2_no_Obs}) with $V$ defined in Fig.~\ref{fig:Harlow} and equation (\ref{eq:Harlow}).
Each term is $\mathcal{O}(1)$, so the variance (and therefore the dimension of the fundamental Hilbert space) is $\mathcal{O}(1)$.

When an observer is included at node $i=1$, that node is not averaged over and the corresponding legs are always acted on by the identity $\mathbb{1}_1$; this gives rise to AAIL's ``observer must stay in their own universe'' rule. As a result, there is a single dominant term in the average given by all nodes being acted on by the identity,
\begin{equation}
    \overline{|\langle\psi|V_{n,Ob}^\dagger V_{n,Ob}|\phi\rangle|^2} = |\langle\psi|\phi\rangle|^2 + \dots
\end{equation}
This is precisely the term that is canceled by $\overline{\langle\psi|V_{n,Ob}^\dagger V_{n,Ob}|\phi\rangle}$ in the calculation of the variance. 
Subleading terms may be characterized by the number of domain walls between mismatching factors of $\mathbb{1}$, $F$, or $M$ acting on each node, remembering that the observer always receives the identity $\mathbb{1}$. Each domain wall contributes a state-independent factor of $1/d_a$ for the maximally entangled leg that is cut. Thus the dominant subleading terms are those with two domain walls, with nodes around the observer being acted on by $\mathbb{1}$ and the rest acted on either all by $F$ or all by $M$. All possible positions for the two domain walls will contribute to the average, differing only by state-dependent contributions. So long as the number of nodes is less than $\mathcal{O}(e^{S_0})$, we will assume these state-dependent contributions don't alter the $1/d_a^2 = e^{-2S_0}$ order of these dominant subleading terms. 

If the state-dependent contributions don't vary much with the position of the cuts, there will be a degeneracy in which cut is dominant in the average. We can lift this degeneracy by allowing each $d_{a_i}$ to be arbitrary but still on the order of $e^{S_0}$. Thus the dominant contribution will be the one that minimizes the bond dimension of the in-plane legs cut by the two domain walls. As an example, the dominant terms in the average of the inner product squared might look like
\begin{align*}
    &\overline{|\langle\psi|V_{n,Ob}^\dagger V_{n,Ob}|\phi\rangle|^2} \\
        &\quad = |\langle\psi|\phi\rangle|^2 + \mathcal{O}(e^{-2S_0}) \Bigg( \tr \Big[ \big( \mathbb{1}_{1} \otimes_{i\neq1} F_i \big) \big( \ket{\phi} \ket{\psi} \bra{\psi} \bra{\phi} \big) \Big] \\
        &\hspace{4cm} + \tr \Big[ \big( \mathbb{1}_{1} \otimes_{i\neq1} M_i \big) \big( \ket{\phi} \ket{\psi} \bra{\psi} \bra{\phi} \big) \Big] \Bigg) + \mathcal{O} \left( e^{-3S_0} \right) \\
        &\quad = |\langle\psi|\phi\rangle|^2 + \mathcal{O}(e^{-2S_0}) \Big( \tr \big[ \tr_{i\neq1} \rho_\psi \cdot \tr_{i\neq1} \rho_\phi \big] + \tr \big[ \rho_\psi^{\intercal_{i\neq1}} \cdot \rho_\phi^{\intercal_{i\neq1}} \big] \Big) + \mathcal{O}\left( e^{-3S_0}\right) \numberthis
\end{align*}
where the cuts have been placed between the observer and the nearest neighboring nodes. Labeling node $i=1$ as $Ob$ and nodes $i\neq1$ as $M$, we find that these are the same leading terms in (\ref{eq:Harlow_IP2_w_Obs_leading}) found with $V_{Ob}$ defined in Fig.~\ref{fig:Harlow_Obs} and equation (\ref{eq:Harlow_Ob}). The first subleading terms are $\mathcal{O}(e^{-2S_0})$, and these are the terms that survive in the variance. Therefore, the variation of the inner product is $\mathcal{O}(e^{-2S_0})$, leading to the fundamental Hilbert space dimension reported in Section \ref{sec:TN} when an observer is included in the tensor network.

\section{Renyi entropy computation of $\dim \mathcal{H}_{\mathrm{fun}}$}
\label{app:renyi}

It is arguably manifest that the tensor networks (TNs) in Fig.\ \ref{fig:TN} have  fundamental Hilbert space dimensions 
\begin{align*}
    \text{No observer:}& \qquad \dim \mathcal{H}_\mathrm{fun} = 1 \\
    \text{Observer:}& \qquad \dim \mathcal{H}_\mathrm{fun} = d_{Ob}e^{2S_0}
\end{align*}
After all, the TN with no observer has no ``boundary'' legs, while the TN with an observer has three ``boundary'' legs, $b_1, a_1, a'_1$, with $d_{a_1} = d_{a'_1} = e^{S_0}$.
In this appendix we explain more concretely how to derive these dimensions from a computation of the second Renyi entropy. 

The idea is to introduce a reference system $R$ that we try our best to entangle with $\mathcal{H}_\mathrm{fun}$.
We will argue that the largest we can possibly make $S_2(R)$ is $0$ in the case of no observer and $\log(d_{b_1}d_{a_1}d_{a'_1})$ in the case that an observer lives at $b_1$.
(Both answers will be accurate only up to $\mathcal{O}(1/d)$ for $d$ the dimension of some Hilbert space factors.)
This will imply the claimed $\dim \mathcal{H}_{\mathrm{fun}}$ because for a bipartite system $\mathcal{H}_{B} \otimes \mathcal{H}_R$ in which $\dim \mathcal{H}_R > \dim \mathcal{H}_B$,
\begin{equation}
    \log \dim \mathcal{H}_{B} = \max_{\ket{\phi} \in \mathcal{H}_{B} \otimes \mathcal{H}_{R}} S_2(R)_{\ket{\phi}}~,
\end{equation}
where $S_2(R)_{\ket{\phi}}$ is the second Renyi entropy of $R$ in the state $\ket{\phi}$. 

We start by considering an arbitrary bulk state $\ket{\psi}_{bR} \equiv \ket{\psi}_{b_1 \dots b_n R}$ and no observer.
Here $R$ is a reference system of arbitrarily large dimension.
Let $V$ denote the tensor network on the left of Fig.~\ref{fig:TN}, with no observer, and let $\rho = V  \ket{\psi}\bra{\psi}_{bR} V^{\dagger}$. 
We will consider the limit where all the bond dimensions $d_i$ are large but distinct.
Using the tools from Appendix \ref{app:orthogonal}, we compute the averaged second Renyi entropy $S_2(R)$ of reference $R$:
\begin{equation}
 \overline{e^{-S_2(R)} } = \overline{ \left( \frac{\tr [(\rho \otimes \rho) F_R ]}{\tr [\rho \otimes \rho]} \right)} ,
\end{equation}
Defining
\begin{align}
  Z_1 &= \tr \left[ (\rho \otimes \rho) F_R \right], \\
  Z_0 &= \tr \left[ \rho \otimes \rho \right],
\end{align}
we can write the averaged second R{\'e}nyi entropy as
\begin{equation}
  S_2(R) = - \overline{\log \frac{Z_1}{Z_{0}}} \simeq - \log \frac{\overline{Z_1}}{\overline{Z_0}} + \cdots ,
\end{equation}
where the $\cdots$ we anticipate to be $\mathcal{O}(1 / d)$, as in the case where the average is over unitary matrices \cite{Hayden:2016cfa}. 
Due to the $F_R$ insertion in $\overline{Z_1}$, its leading order contribution will be the term with swap operators on all nodes. Similarly, we will have identity operators on all nodes for $\overline{Z_0}$. Hence $\overline{Z_1} = \overline{Z_0} = 1$ up to $\mathcal{O}(1 / d)$ corrections, giving $S_2(R) = \mathcal{O}(1/d)$.

Now we consider the case where an observer is present at node $1$. This puts an identity operator at the observer's node, forcing some combination of identity and swap operators acting at different nodes on the network and reference when we compute $\overline{Z_1}$. The computation of $\overline{Z_0}$ is the same as in the no-observer case. The leading order contribution to $\overline{Z_1}$ will either have identities at all nodes, or will cut the tensor network into two connected segments of nodes. The segment with identity operators will always include the observer, and the bulk legs attached to the nodes $\{1, j_1, \dots, j_k\}$ in this segment will contribute a term $S_2(b_1 b_{j_1} \dots b_{j_k})_{\ket{\psi}_{bR}}$ to $S_2(R)$. The other segment, if it exists, will receive swap operators. This means that there will be two in-plane legs, which we will call $a$ and $\bar{a}$, with an identity on one end and a swap on the other, contributing a term $\log (d_{a} d_{\bar{a}})$ to $S_2(R)$.

If we approximate $\overline{Z_1}$ by this leading contribution, we obtain an answer called the ``quantum minimal surface'' value of the Renyi entropy, $S_2^{\mathrm{QMS}}(R)$, given by
\begin{equation}
  S^{\mathrm{QMS}}_2(R)_{V\ket{\psi}} = \min \left\{ \log (d_{a} d_{\bar{a}}) + S_2(b_1 b_{j_1} \dots b_{j_k})_{\ket{\psi}_{bR}} \right\} + \mathcal{O}(1/d)~,
\end{equation}
where the minimization is over the cut, i.e. which in-plane legs are $a$ and $\bar{a}$. 
Of course, this approximation of $\overline{Z}_1$ is not always valid, and for some $\ket{\psi}_{bR}$ can be very wrong \cite{Akers:2020pmf}.
Nonetheless, for any $\ket{\psi}_{bR}$ it holds that
\begin{equation}
    S^{\mathrm{QMS}}_2(R)_{V\ket{\psi}} \ge S_2(R)_{V\ket{\psi}}~.
\end{equation}
This follows because the approximation removed positive contributions from $\overline{Z}_1$, increasing $-\log \overline{Z}_1$. 
The inequality can be approximately saturated, and indeed is approximately saturated when maximizing both sides over the choice of $\ket{\psi}_{bR}$:
\begin{equation}
    \max_{\ket{\psi}} S^{\mathrm{QMS}}_2(R)_{V\ket{\psi}} = \max_{\ket{\psi}} S_2(R)_{V\ket{\psi}} + \mathcal{O}(1/d)~.
\end{equation}
This is because in the case that $\ket{\psi}_{bR} = \ket{\mathrm{MAX}}_{bR}$, the quantum minimal surface formula is indeed valid:
\begin{equation}
 S_2(R)_{V\ket{\mathrm{MAX}}} = S^{\mathrm{QMS}}_2(R)_{V\ket{\mathrm{MAX}}} + \mathcal{O}(1/d) = \min \left\{ \log (d_{a} d_{\bar{a}} d_{b_1} d_{b_{j_{1}}} \dots d_{b_{j_{k}}} ) \right\} + \mathcal{O}(1/d),
\end{equation}
and furthermore $S_2^{\mathrm{QMS}}(R)$ is maximized in this case,
\begin{equation}
    \max_{\ket{\psi}} S_2^{\mathrm{QMS}}(R)_{V \ket{\psi}} = S_2^{\mathrm{QMS}}(R)_{V \ket{\mathrm{MAX}}}.
\end{equation}
Therefore, 
\begin{equation}
     S_2^{\mathrm{QMS}}(R)_{V \ket{\mathrm{MAX}}} = \max_{\ket{\psi}} S_2^{\mathrm{QMS}}(R)_{V \ket{\psi}} \approx \max_{\ket{\psi}} S_2(R)_{V \ket{\psi}} = \log (\dim \mathcal{H}_{\mathrm{fun}})~.
\end{equation}
This implies that up to factors of order $\exp(1/d)$, 
\begin{align}
  \text{No observer:}& \qquad \dim \mathcal{H}_{\mathrm{fun}} = 1 \nonumber \\
  \text{Observer:}& \qquad \dim \mathcal{H}_{\mathrm{fun}} = \min d_{a} d_{\bar{a}} d_{b_1} d_{b_{j_{1}}} \dots d_{b_{j_{k}}}.
\end{align}
Again, the minimization is over the location of the cut dividing $b_1,b_{j_1},...b_{j_k}$ from the other bulk legs. 
This is almost what we set out to show. 
In the observer case, if we would like this to model semiclassical JT gravity, then we should make specific choices for the dimensions of each factor.
The in-plane legs should have $d_a = d_{\bar{a}} = e^{S_0}$, and the $b_i$ should have some large dimension, modeling the quantum field theory coupled to JT.
In that case, the minimization excludes all $b_i$ except $b_1$, whose dimension we write as $d_{b_1} = d_{Ob}$, leaving 
\begin{align}
  \text{Observer:}& \qquad \dim \mathcal{H}_{\mathrm{fun}} = d_{a} d_{\bar{a}} d_{b_1} = e^{2S_0} d_{Ob}~.
\end{align}

\bibliographystyle{jhep}
\bibliography{mybib2025}

\end{document}